# The Dirac Equation in the Model With a Maximal Mass

*V.P. Neznamov*

*E-mail address: neznamov@vniief.ru*

*RFNC-VNIIEF, 607188, Sarov, N.Novgorod region, Russia*

## Abstract

In paper within the model with a maximal mass M and with use of anti de Sitter space is considered the Dirac equation properties for a fermion of mass *m* on the mass surface ($p_5 = \pm\sqrt{M^2 - m^2}$).

The paper shows that free Hamiltonian and Hamiltonian with interaction are pseudo-Hermitian.

If we accept the appropriate rules of use of pseudo-Hermitian Hamiltonians, we can construct the consistent quantum theory with final results obtained using the ordinary Dirac equation.

## 1. INTRODUCTION

V.G.Kadyshevsky and his colleagues in their papers [2] – [9] developed the M.A. Markov's idea about existence of a maximal elementary particle mass M [1]. In these papers the existence of mass M has been understood as a new fundamental principle of Nature similar to the relativistic and quantum postulates, which are put in the grounds of quantum field theory.

The condition of finiteness of elementary particle mass spectrum is expressed by introduction of relation

$$m \leq M, \qquad (1)$$

where mass parameter M is a new physical constant. In the paper [1] by M.A.Markov,

$$M \cong m_{planck} = 10^{19} GeV.$$

In the papers [2] – [9], a new concept of a local quantum theory has been developed on the ground of relation (1) with construction of corresponding Lagrangians for boson and fermion fields. Relation (1) is satisfied by introduction of anti de Sitter space just as for satisfaction of relativistic conditions in due time was necessary the transition from a 3D space to a 4D Minkovsky space.

An analogue of the relativistic relation between energy and momentum of particle with fulfillment of relation (1) in anti de Sitter momentum space may be written as

$$p_0^2 - \mathbf{p}^2 + p_5^2 = M^2. \qquad (2)$$

The relation $p_0^2 - \mathbf{p}^2 = m^2$ is satisfied for particles of mass $m$ on the mass surface and apparently satisfaction of equality

$$p_5^2 = M^2 - m^2. \qquad (3)$$

As it has been noted in [9], the introduction of mass parameter M means an existence of a fundamental length $l = \hbar / Mc$. In recent years the interest in the introduction of a minimal length has grown in connection with the theoretical studies in quantum gravitation theory and theory of strings (see, for example, [10 – 12]).

This paper investigates the properties of the Dirac equation introduced in [9] for fermions on the mass surface with a maximal mass M.

## 2. THE DIRAC EQUATION WITH A MAXIMAL MASS M

First, consider the free motion of a Dirac particle of mass $m$.

According to [9] equation (2) on mass surface $p_0^2 - \mathbf{p}^2 = m^2$ may be written in the form

$$(p_5 + M \cos \mu)(p_5 - M \cos \mu) = 0, \qquad (4)$$

where $\cos \mu = \sqrt{1 - \dfrac{m^2}{M^2}}$.

Each of the factors in (4) can be represented as

$$-2M(p_5 + M \cos \mu) = (\gamma^0 p_0 - \boldsymbol{\gamma}\mathbf{p} - \gamma^5(p_5 + M) - 2M \sin \frac{\mu}{2})(-\gamma^0 p_0 + \boldsymbol{\gamma}\mathbf{p} + \gamma^5(p_5 + M) - 2M \sin \frac{\mu}{2}); \qquad (5)$$

$$2M(p_5 - M \cos \mu) = (\gamma^0 p_0 - \boldsymbol{\gamma}\mathbf{p} - \gamma^5(p_5 - M) + 2M \sin \frac{\mu}{2})(-\gamma^0 p_0 + \boldsymbol{\gamma}\mathbf{p} + \gamma^5(p_5 - M) + 2M \sin \frac{\mu}{2}). \qquad (6)$$

In expression (5) $p_5 = -\sqrt{M^2 - m^2} = -M \cos \mu$; in expression (6) $p_5 = \sqrt{M^2 - m^2} = M \cos \mu$.

Consider expressions (5), (6) as operators in configuration space $(x, x_5) = (x^0, \mathbf{x}, x_5)$ and apply these to Dirac bi-spinor $\psi(x, x_5)$ to obtain the following four equations:

$$\begin{cases} p_0 \psi_1(x,x_5) = (\boldsymbol{\alpha}\mathbf{p} + \beta\gamma^5 m_1 + \beta m_2)\psi_1(x,x_5) \\ p_0 \psi_2(x,x_5) = (\boldsymbol{\alpha}\mathbf{p} + \beta\gamma^5 m_1 - \beta m_2)\psi_2(x,x_5) \end{cases} \quad p_5 = -\sqrt{M^2 - m^2};$$

$$\begin{cases} p_0 \psi_3(x,x_5) = (\boldsymbol{\alpha}\mathbf{p} - \beta\gamma^5 m_1 + \beta m_2)\psi_3(x,x_5) \\ p_0 \psi_4(x,x_5) = (\boldsymbol{\alpha}\mathbf{p} - \beta\gamma^5 m_1 - \beta m_2)\psi_4(x,x_5) \end{cases} \quad p_5 = \sqrt{M^2 - m^2}. \qquad (7)$$

In equations (5) – (7) and below, $\hbar = c = 1$; $\gamma^0 = \beta$; $\gamma^i = \beta\alpha^i$; $\gamma^5 = \gamma^0\gamma^1\gamma^2\gamma^3$ are four-dimensional Dirac matrices; $m_1 = 2M \sin^2 \frac{\mu}{2}$; $m_2 = 2M \sin\frac{\mu}{2}$; $m_2^2 - m_1^2 = m^2$; for a case $m \ll M$, $m_1 \approx \frac{m^2}{2M}$, $m_2 \approx m(1 + \frac{1}{8}\frac{m^2}{M^2})$.

Equations (7) differ from each other only in their signs before the terms with $m_1$ and $m_2$. As for their physical consequences, equations (7) are equivalent to each other similarly to the ordinary Dirac equations differing in their signs before the mass term.

Expression (4) also allows decomposition of another kind [9], which may be written briefly as

$$\pm 2M(p_5 \mp M\cos\mu) = (-\gamma^0 p_0 + \boldsymbol{\gamma}\mathbf{p} - \gamma^5(p_5 \pm M) \pm 2M\cos\frac{\mu}{2})(-\gamma^0 p_0 + \boldsymbol{\gamma}\mathbf{p} - \gamma^5(p_5 \pm M) \mp 2M\cos\frac{\mu}{2}). \qquad (8)$$

In expression (8), $p_5 = +\sqrt{M^2 - m^2}$ for the upper signs and $p_5 = -\sqrt{M^2 - m^2}$ for the lower signs.

Similarly to (7), one may write four equivalent equations differing in their signs before the mass terms

$$\begin{cases} p_0 \psi_5(x,x_5) = (\boldsymbol{\alpha}\mathbf{p} + \beta\gamma^5 m_3 + \beta m_4)\psi_5(x,x_5) \\ p_0 \psi_6(x,x_5) = (\boldsymbol{\alpha}\mathbf{p} + \beta\gamma^5 m_3 - \beta m_4)\psi_6(x,x_5) \end{cases} \quad p_5 = -\sqrt{M^2 - m^2};$$

$$\begin{cases} p_0 \psi_7(x,x_5) = (\boldsymbol{\alpha}\mathbf{p} - \beta\gamma^5 m_3 + \beta m_4)\psi_7(x,x_5) \\ p_0 \psi_8(x,x_5) = (\boldsymbol{\alpha}\mathbf{p} - \beta\gamma^5 m_3 - \beta m_4)\psi_8(x,x_5) \end{cases} \quad p_5 = \sqrt{M^2 - m^2}. \qquad (9)$$

In equations (9), $m_3 = 2M\cos^2\frac{\mu}{2}$; $m_4 = 2M\cos\frac{\mu}{2}$.

As opposed to the values of $m_1$, $m_2$ in (7), for a case of $m \ll M$ the values of masses $m_3$, $m_4$ are large ($m_3 \approx 2M(1 - \frac{1}{4}\frac{m^2}{M^2})$; $m_4 \approx 2M(1 - \frac{1}{8}\frac{m^2}{M^2})$, but $m_4^2 - m_3^2 = m^2$).

In equations (7), (9) for fermions on the mass surface there are not the operators, which act on the coordinate $x_5$ and $x_5$ can be taken equal to zero.

## 3. THE HERMITIAN PROPERTIES OF DIRAC HAMILTONIAN WITH THE MAXIMAL MASS M

### 3.1 Free motion of a Dirac particle of mass *m*

Any of the equivalent Hamiltonians from equations (7), (9) may be taken for further consideration.

Let us have

$$H_0 = \boldsymbol{\alpha}\mathbf{p} + \beta\gamma^5 m_1 + \beta m_2. \tag{10}$$

It is clear that expression (10) is non-Hermitian ($H_0 \neq H_0^\dagger$) because of the term with $m_1$. However, $H_0^2 = E^2 = m^2 + \mathbf{p}^2$ and, apparently, solutions of the Dirac equation with Hamiltonian (10) will be plane waves with continuous energy spectrum, $E = \pm\sqrt{m^2 + \mathbf{p}^2}$.

Analyze Hamiltonian (10) from non-Hermitian quantum mechanics point of view showing its rapid development in recent years. The fundamental doctrine of quantum mechanic was that Hamiltonian and all physical observables should be represented by Hermitian matrices. More a decade ago, Bender et al [13] showed that this principle could be soften to some extent. The Hermiticity in the canonic quantum-mechanics formulation provides the real spectrum of eigenvalues of the operators under consideration. The authors of [13] demonstrated that the Hermiticity is sufficient, but not necessary for these purposes. The reality of eigenvalues and consistent quantum-mechanics description may be formulated with a number of more general requirements in the absence of Hermiticity.

Operators may be quasi-Hermitian [14], or pseudo-Hermitian [15], and/or PT-symmetric [16].

A.Mostafazadeh identified the necessary and sufficient requirements of reality of eigenvalues for pseudo-Hermitian and PT-symmetric Hamiltonians and formalized the use of these Hamiltonians in his papers [15], [17].

According to [15] and [17] define Hermitian operator $\rho$, which transforms Hamiltonian (10) by means of invertible transformation to the Hermitian-conjugated one

$$\rho H_0 \rho^{-1} = H_0^\dagger. \tag{11}$$

The desired operator $\rho$ may be represented as

$$\rho = 1 + \frac{m_1^2}{E(E+m_2)} + \frac{\gamma^5 m_1}{E} + \frac{\gamma^5 m_1 \boldsymbol{\beta\alpha p}}{E(E+m_2)}; \tag{12}$$

$$\rho^{-1} = 1 + \frac{m_1^2}{E(E+m_2)} - \frac{\gamma^5 m_1}{E} - \frac{\gamma^5 m_1 \boldsymbol{\beta\alpha p}}{E(E+m_2)}. \tag{13}$$

Expressions (12), (13) provide the fulfillment of pseudo-Hermiticity conditions (11). The inner product in Hilbert space is written, in this case, with the weight of operator $\rho$

$$<\phi/\psi>_\rho = <\phi/\rho\psi>. \tag{14}$$

If operator $\rho$ may be represented as a product of two operators,

$$\rho = \eta^\dagger \eta, \tag{15}$$

Hamiltonian (10) can be transformed to its Hermitian form, i.e. with preservation of eigenvalues.

$$\eta H_0 \eta^{-1} = H_{FW} = H_{FW}^\dagger \tag{16}$$

For operator $\rho$ of the form (12) the expressions for operators $\eta$, $\eta^{-1}$ look like

$$\eta = \sqrt{\frac{E+m_2}{2E}}(1+\frac{1}{E+m_2}(\beta\alpha\mathbf{p}+\gamma^5 m_1)); \tag{17}$$

$$\eta^{-1} = \sqrt{\frac{E+m_2}{2E}}(1+\frac{1}{E+m_2}(\alpha\mathbf{p}\beta-\gamma^5 m_1)). \tag{18}$$

Taking into account (17), (18) expression (16) equals

$$\eta(\alpha\mathbf{p}+\beta\gamma^5 m_1+\beta m_2)\eta^{-1} = \beta E = \beta\sqrt{m^2+\mathbf{p}^2}. \tag{19}$$

Expression (19) is actually the free Foldy-Wouthuysen Hamiltonian [18] with the spectrum of eigenvalues $\pm\sqrt{m^2+\mathbf{p}^2}$.

Expressions (17), (18) are the nonunitary Foldy-Wouthuysen transformation operators concerning Hamiltonian (10).

Thus, we have demonstrated that Hamiltonian $H_0$ is pseudo-Hermitian and has the same spectrum of eigenvalues, as the free Foldy-Wouthuysen Hamiltonian.

Solutions of equations (7), (9) are the plane waves with positive and negative energy:

$$\psi_D^{(+)}(x,s) = U_s e^{-ipx}; \quad \psi_D^{(-)}(x,s) = V_s e^{ipx},$$

$$p_0 = E = (m^2+\mathbf{p}^2)^{\frac{1}{2}}. \tag{20}$$

For Hamiltonian (10),

$$U_s = \sqrt{\frac{E+m_2}{2E}}\begin{pmatrix} \varphi_s \\ \frac{\sigma\mathbf{p}-m_1}{E+m_2}\varphi_s \end{pmatrix}; \quad V_s = \sqrt{\frac{E+m_2}{2E}}\begin{pmatrix} \frac{\sigma\mathbf{p}-m_1}{E+m_2}\chi_s \\ \chi_s \end{pmatrix}. \tag{21}$$

In (21) $\mathbf{p}$ and $E$ are the momentum and energy operators for a particle of mass $m$; $\varphi_s$ and $\chi_s$ are the 2-component normalized Pauli spin functions.

The following relations of orthonormality and completeness are valid for $U_s$ and $V_s$ taking into account the weight operators $\rho$ (12):

$$U_s^\dagger \rho U_{s'} = V_s^\dagger \rho V_{s'} = \delta_{ss'}; \quad U_s^\dagger \rho V_{s'} = V_s^\dagger \rho U_{s'} = 0;$$

$$\sum_s (U_s)_\alpha (U_s^\dagger)_\beta \rho_{\beta\delta} = \frac{1}{2}(1+\frac{H_0}{E})_{\alpha\delta}; \tag{22}$$

$$\sum_s (V_s)_\alpha (V_s^\dagger)_\beta \rho_{\beta\delta} = \frac{1}{2}(1-\frac{H_0}{E})_{\alpha\delta}.$$

In expressions (20) – (22), $\alpha$, $\beta$, and $\delta$ are used for spinor indexes and $s$, $s'$ is used for spin indexes. Further the summation symbol and indexes themselves may not be shown when summation with respect to spinor indexes.

## 3.2 Motion of the interactive Dirac particle

In the presence of a gauge vector Abelian field $B^\mu(x)$ Hamiltonian (10) may be written as

$$H_D = \boldsymbol{\alpha}\mathbf{p} + \beta\gamma^5 m_1 + q\alpha_\mu B^\mu \qquad (23)$$

In expression (23), $\alpha^\mu = \begin{cases} 1 & \mu = 0 \\ \alpha^i & \mu = i = 1,2,3 \end{cases}$; $q$ is a coupling constant.

The Abelian case of field $B^\mu(x)$ is considered for simplicity. The conclusions made in this paper are not varied for a general case of a Dirac particle interacts with a non-Abelian gauge field.

The form of operators $\rho$ and $\eta$ used to reduce Hamiltonian (23) to its Hermitian form with the preservation of eigenvalues has not been found by present time.

However, one can demonstrate that Hamiltonian (23) together with the Hamiltonian differing from (23) in the sign standing before the non-Hermitian term of mass $m_1$ are both pseudo-Hermitian and, hence, they allow using them within the formalism [15, 17].

Indeed, introduce an eight-component spinor $\phi(x)$ with four upper components being the solutions of the Dirac equation with Hamiltonian (23) and four lower components being the solutions of the Dirac equation with Hamiltonian differing from (23) in the sign before the term with mass $m_1$. Introduce also isotopic matrices $\tau_3 = \begin{pmatrix} I & 0 \\ 0 & -I \end{pmatrix}$ and $\tau_1 = \begin{pmatrix} 0 & I \\ I & 0 \end{pmatrix}$, which are effective in the isotopic space of the four upper and four lower components of spinor $\phi(x)$.

Then, the two Dirac equations (7) with the introduced interaction with field $B^\mu(x)$ may be written as

$$\begin{cases} p_0\psi_1(x) = (\boldsymbol{\alpha}\mathbf{p} + \beta\gamma^5 m_1 + \beta m_2 + q\alpha_\mu B^\mu(x))\psi_1(x); \quad p_5 = -\sqrt{M^2 - m^2} \\ p_0\psi_3(x) = (\boldsymbol{\alpha}\mathbf{p} - \beta\gamma^5 m_1 + \beta m_2 + q\alpha_\mu B^\mu(x))\psi_3(x); \quad p_5 = \sqrt{M^2 - m^2} \end{cases} ; \qquad (24)$$

$$\phi(x) = \begin{pmatrix} \psi_2(x) \\ \psi_3(x) \end{pmatrix}; \qquad (25)$$

$$p_0\phi(x) = (\boldsymbol{\alpha}\mathbf{p} + \tau_3\beta\gamma^5 m_1 + \beta m_2 + q\alpha_\mu B^\mu(x))\phi(x). \qquad (26)$$

In expressions (23) – (26) as in the previous subsection the coordinate $x_5$ can be taken equal to zero [9].

As one may see from (24), equation (26) has two branches of possible values of $p_5 = \pm\sqrt{M^2 - m^2}$.

It is clear from equation (26) that matrix $\tau_1 = \tau_1^{-1}$ plays the role of operator $\rho$ in this case

$$\tau_1 H_\phi \tau_1 = \tau_1(\boldsymbol{\alpha}\mathbf{p} + \tau_3\beta\gamma^5 m_1 + \beta m_2 + q\alpha_\mu B^\mu(x))\tau_1 = H_\phi^\dagger. \qquad (27)$$

Equation (23) allows concluding that Hamiltonian $H_\phi$ in equation (26) is pseudo-Hermitian.

Equation (26) together with its Hamiltonian can be transformed to the Foldy-Wouthuysen representation. In this case we use the formalism earlier developed by author for the standard Dirac equation [20] and the generalized Hermitian conjugation conditions for any operators $(L^\dagger)_{gen} = \tau_1 L^\dagger \tau_1$, which follows from pseudo-Hermitian Hamiltonian (26).

The Hamiltonian of the equation (26) in the Foldy-Wouthuysen representation is Hermitian with the real spectrum of eigenvalues.

All the previous conclusions for the ordinary, modified and isotopic Foldy-Wouthuysen representation [20, 21] are also valid for the case under consideration. In particular, the Foldy-Wouthuysen transformation matrices for the free motion coincide with matrices $\eta$, $\eta^{-1}$ obtained in subsection 3.1. Matrix $\eta^{-1} = \tau_1 \eta^\dagger \tau_1 = ((U_0)_{FW}^\dagger)_{gen}$ should be used for $\eta^{-1}$.

The basic functions of the free motion with $B^\mu = 0$ for equation (26) correspond to expressions (20), (21) with selection of an appropriate sign before $m_1$. According to (26), there are two positive-energy and two negative-energy solutions:

$$\phi_1^{(+)}(x,s) = U_s(T_3 = +1/2)e^{-ipx} = \begin{pmatrix} U_s(m_1) \\ 0 \end{pmatrix} e^{-ipx};$$

$$\phi_2^{(+)}(x,s) = U_s(T_3 = -1/2)e^{-ipx} = \begin{pmatrix} 0 \\ U_s(-m_1) \end{pmatrix} e^{-ipx}; \qquad (28)$$

$$\phi_1^{(-)}(x,s) = V_s(T_3 = +1/2)e^{ipx} = \begin{pmatrix} V_s(m_1) \\ 0 \end{pmatrix} e^{ipx};$$

$$\phi_2^{(-)}(x,s) = V_s(T_3 = -1/2)e^{ipx} = \begin{pmatrix} 0 \\ V_s(-m_1) \end{pmatrix} e^{ipx}.$$

In the expression (28) $T_3$ is the third component of isotopic spin.

The orthonormality and completeness relations for an eight-component functions $U_s(T_3 = \pm 1/2)$, $V_s(T_3 = \pm 1/2)$ have the following view (below $H_0$ is Hamiltionian of equation (26) with $B^\mu = 0$).

$$U_s^\dagger(T_3 = \mp 1/2)\tau_1 U_{s'}(T_3 = \pm 1/2) = V_s^\dagger(T_3 = \mp 1/2)\tau_1 V_{s'}(T_3 = \pm 1/2) = \delta_{ss'};$$

$$U_s^\dagger(T_3 = \mp 1/2)\tau_1 V_{s'}(T_3 = \pm 1/2) = V_s^\dagger(T_3 = \mp 1/2)\tau_1 U_{s'}(T_3 = \pm 1/2) = 0;$$

$$\sum_s (U_s(T_3 = \pm 1/2))_\alpha (U_s^\dagger(T_3 = \mp 1/2))_\beta (\tau_1)_{\beta\gamma} = \left[\frac{1}{2}\left(1 + \frac{H_0}{E}\right)\frac{1}{2}(1 \pm \tau_3)\right]_{\alpha\gamma};$$

$$\sum_s (V_s(T_3 = \pm 1/2))_\alpha (V_s^\dagger(T_3 = \mp 1/2))_\beta (\tau_1)_{\beta\gamma} = \left[\frac{1}{2}\left(1 - \frac{H_0}{E}\right)\frac{1}{2}(1 \pm \tau_3)\right]_{\alpha\gamma}.$$

Similar to equations (7) and (9), equation (26) may be written in the relativistic covariant form

$$(\hat{p} - \tau_3 \gamma^5 m_1 - m_2 - q\gamma_\mu B^\mu)\phi(x) = 0. \qquad (29)$$

In equation (29) $\hat{p} = \gamma^0 p^0 - \boldsymbol{\gamma}\mathbf{p}$.

We can neglect the term with $m_1$ in (29) and (7), if $m \ll M$, and obtain the ordinary Dirac equation. For equation (9), the both mass terms with $m_3$ and $m_4$ have to be necessarily taken into account, however, on the mass surface $m_4^2 - m_3^2 = m^2$ in any case.

Dirac functions $u(p,s)$, $v(p,s)$ [19] generalized to eight-component case and with an appropriate value $T_3 = \pm 1/2$ will be analogues of basic functions (28) for equation (29). The following relations are valid for these functions:

$$(\hat{p} - \tau_3 \gamma^5 m_1 - m_2)u(p,s,T_3) = 0; \quad \bar{u}\tau_1(\hat{p} - \tau_3 \gamma^5 m_1 - m_2) = 0;$$

$$(\hat{p} + \tau_3 \gamma^5 m_1 + m_2)v(p,s,T_3) = 0; \quad \bar{v}\tau_1(\hat{p} + \tau_3 \gamma^5 m_1 + m_2) = 0; \qquad (30)$$

$$\bar{u}_s(T_3 = \mp 1/2)\tau_1 u_{s'}(T_3 = \pm 1/2) = \bar{v}_s(T_3 = \mp 1/2)\tau_1 v_{s'}(T_3 = \pm 1/2) = \delta_{ss'}; \quad (31)$$

$$\bar{u}_s(T_3 = \mp 1/2)\tau_1 v_{s'}(T_3 = \pm 1/2) = \bar{v}_s(T_3 = \mp 1/2)\tau_1 u_{s'}(T_3 = \pm 1/2) = 0.$$

The projection operators for the input fermion lines in Feynman diagrams are:

$$\Lambda_u^{(+)}(p) = \frac{1}{2(\tau_3\gamma^5 m_1 + m_2)}(\hat{p} + \tau_3\gamma^5 m_1 + m_2) = \frac{(m_2 - \tau_3\gamma^5 m_1)\hat{p} + m^2}{2m^2};$$

$$\Lambda_u^{(+)} u = u; \quad (\Lambda_u^{(+)})^2 u = \Lambda_u^{(+)} u; \quad (32)$$

$$\Lambda_v^{(-)}(p) = -\frac{(m_2 - \tau_3\gamma^5 m_1)\hat{p} - m^2}{2m^2}; \quad \Lambda_v^{(-)} v = v; \quad (\Lambda_v^{(-)})^2 v = \Lambda_v^{(-)} v.$$

The projection operators for the output fermion lines look like

$$\Lambda_{\bar{u}}^{(+)}(p) = \frac{\hat{p}(m_2 - \tau_3\gamma^5 m_1) + m^2}{2m^2}; \quad \bar{u}\tau_1\Lambda_{\bar{u}}^{(+)} = \bar{u}\tau_1; \quad \bar{u}\tau_1(\Lambda_{\bar{u}}^{(+)})^2 = \bar{u}\tau_1\Lambda_{\bar{u}}^{(+)};$$

$$\Lambda_{\bar{v}}^{(-)}(p) = -\frac{\hat{p}(m_2 - \tau_3\gamma^5 m_1) - m^2}{2m^2}; \quad \bar{v}\tau_1\Lambda_{\bar{v}}^{(-)} = \bar{v}\tau_1; \quad \bar{v}\tau_1(\Lambda_{\bar{v}}^{(-)})^2 = \bar{v}\tau_1\Lambda_{\bar{v}}^{(-)}. \quad (33)$$

The completeness relations for the input and output fermion lines are:

$$\sum_s \left(u_s(T_3 = \pm 1/2)\right)_\alpha \left(\bar{u}_s(T_3 = \mp 1/2)\right)_\beta (\tau_1)_{\beta\gamma} = \left(\Lambda_u^{(+)} 1/2(1 \pm \tau_3)\right)_{\alpha\gamma};$$

$$\sum_s \left(v_s(T_3 = \pm 1/2)\right)_\alpha \left(\bar{v}_s(T_3 = \mp 1/2)\right)_\beta (\tau_1)_{\beta\gamma} = \left(\Lambda_v^{(-)} 1/2(1 \pm \tau_3)\right)_{\alpha\gamma}. \quad (34)$$

In expressions (30) – (34), $\bar{u}$ and $\bar{v}$ denote, as usually, the Hermitian conjugation of spinors $u$, $v$ with the simultaneous multiplication by matrix $\gamma^0$.

## 4. DISCUSSION OF RESULTS

The investigations given in the previous sections show that the Dirac Hamiltonians of a particle of mass $m$ with maximal mass parameter $M$ are pseudo-Hermitian. It is valid with or without of interactions with the gauge boson fields for the case under consideration $p_5 = \pm\sqrt{M^2 - m^2}$. According to the papers [15], [16] and [17] pseudo-Hermitian operators have either real eigenvalues, or pairs of complex conjugate eigenvalues.

For the free motion, we can reduce the pseudo-Hermitian Hamiltonians of equations (7), (9) to the Hermitian form and, thereby, demonstrate that they have real eigenvalues.

In the presence of interaction the combination of Dirac equations with different signs before the non-Hermitian term with mass $m_1$ lets also transform the obtained pseudo-Hermitian Hamiltonians to the Hermititian form by means of the transition to Foldy-Wouthuysen representation.

Consider the simplest quantum electrodynamics problem of electron scattering in the Coulomb field of a point nucleus of charge $-Ze > 0$ to illustrate the possibility of application of Hamiltonian from equation (26) to solve the quantum field theory's problems.

In this case, the term with interaction $q\alpha_\mu B^\mu$ in (26) is

$$-eA_0(x) = \frac{Ze^2}{4\pi|\mathbf{x}|}. \quad (35)$$

In the standard quantum electrodynamics case (see, for example, [19]), the differential electron scattering cross-section is

$$\frac{d\sigma}{d\Omega} = \frac{2Z^2\alpha^2 m^2}{|\mathbf{q}|^4} \sum_{\pm s_f, s_i} \left| \bar{u}(p_f, s_f) \gamma^0 u(p_i, s_i) \right|^2 =$$
$$= \frac{2Z^2\alpha^2 m^2}{|\mathbf{q}|^4} Sp\left[ \gamma^0 \frac{(\hat{p}_i + m)}{2m} \gamma^0 \frac{(\hat{p}_f + m)}{2m} \right] = \frac{Z^2\alpha^2}{2|\mathbf{q}|^4} Sp(\gamma^0 \hat{p}_i \gamma^0 \hat{p}_f + m^2). \quad (36)$$

In (36) $\mathbf{q} = \mathbf{p}_f - \mathbf{p}_i$ describes the variation of the scattering electron's momentum; $\alpha$ is the thin structure's constant; $p_i$, $s_i$, $p_f$, $s_f$ are the initial (final) four momentum and spin of electron.

In our case, we can obtain the following equation using equations (26), (29) and relations (30) – (34)

$$\frac{d\sigma}{d\Omega} = \frac{2Z^2\alpha^2 m^2}{|\mathbf{q}|^4} Sp\left[ \gamma^0 \frac{(m_2 - \tau_3 \gamma^5 m_1)\hat{p}_i + m^2}{2m^2} \frac{1}{2}(I \pm \tau_3)\gamma^0 \cdot \right.$$
$$\left. \cdot \frac{\hat{p}_f(m_2 - \tau_3 \gamma^5 m_1) + m^2}{2m^2} \frac{1}{2}(I \pm \tau_3) \right]. \quad (37)$$

The matrix trace in expression (37) should be calculated using the eight-component structure of matrices. However, projection operators $1/2(I + \tau_3)$ actually reduce this operation to finding the value of $Sp$ for the four-component matrices.

Taking into account this remark, it becomes clear that expressions (37) and (36) are coincident.

Apparently, we can define other effects of the quantum field theory with Hamiltonian (26), when a fermion is on the mass surface $(p_5 = \pm\sqrt{M_2 + m^2})$.

In contrast to the standard approach, parameter $\tau_3 \gamma^5 m_1 + m_2$ is used instead of mass $m$, as well as relations (30) – (34). The final and standard results are coincident.

The case of fermions and bosons located outside the mass surface requires a special approach and will be considered in the next paper.

Thus, we may conclude that the introduction to theory of a maximal mass $M$ for a fermion on the mass surface $(p_5 = \pm\sqrt{M^2 - m^2})$ does not affect the final results earlier obtained within the standard quantum theory in spite of the initial non-Hermiticity of the Dirac Hamiltonian.

**Acknowledgements.** The author would like to thank V.G.Kadyshevsky for the useful discussions, advices and criticism.